\documentclass[a4paper, 10pt, conference]{ieeeconf}      

\IEEEoverridecommandlockouts                              

\title{\LARGE \bf
	New vehicle-actuated access algorithms \\for intersections close to oversaturation
}
\usepackage{theorem,amsmath,amssymb}
\newtheorem{theorem}{Theorem}
\newtheorem{remark}[theorem]{Remark}

\author{Rik W. Timmerman$^{1}$ and Marko A.A. Boon$^{1}$
	\thanks{This work was supported by NWO under grant 438-13-206.}
	\thanks{$^{1}$ Marko A.A. Boon and Rik W. Timmerman are with the Department of Mathematics and Computer Science,
		Eindhoven University of Technology, 5600 MB Eindhoven, The Netherlands.
		{\tt\small r.w.timmerman@tue.nl}}%
}

\usepackage{url,color,graphicx,float}

\begin{document}

	\maketitle
	\thispagestyle{empty}
	\pagestyle{empty}

	\begin{abstract}
		Optimal traffic-light settings are generally hard to obtain, certainly for actuated access control of an intersection. Typically, computationally expensive (microscopic) simulations or complicated optimization schemes are required to find those optimal settings. Based on recent developments regarding the Fixed-Cycle Traffic-Light (FCTL) queue, we propose a new allocation scheme for actuated access control at isolated intersections. Our scheme allows for very easy computations and an intuitive explanation, which are both based on the Central Limit Theorem, a key tool in probability theory. The desirable properties observed for the FCTL queue can also be achieved in case of actuated access control of an intersection. We show this by means of simulation of the underlying queueing model, because the actuated control defies an exact analysis so far. We also observe the same type of results employing the microscopic traffic simulator SUMO. The intuition and insights obtained might also open up new possibilities for obtaining optimal actuated access control for a network of intersections.
	\end{abstract}

	\section{INTRODUCTION} \label{sec:intro}
	
	Vehicle-actuated access control of intersections is a key tool in keeping congestion at a (relatively) low level in urban areas. Intersections tend to increase the amount of congestion, because they reduce the capacity of the roads which are connected to the intersection. Even though the current state-of-the-art control systems for intersections are smart and adaptive, many cities experience a lot of congestion, as recent statistics of TomTom indicate \cite{tomtom2018}. There are more than a hundred cities where a trip at an arbitrary moment of the day has an increased duration of more than 30\% compared to free flow. This congestion negatively impacts safety and the environment, causes a lot of stress and, obviously, accounts for much lost time.
	
	Control mechanisms such as semi-actuated or fully actuated control are able to adapt to (time-)varying circumstances. Commonly, as a control parameter, a maximum length of the cycle is introduced and hence each lane receives a limited amount of time for vehicle crossing, reducing the probability of excessive waiting times for e.g. lanes on which only few vehicles drive. However, the queueing models underneath such traffic-light settings are not well understood even in the most basic case of an isolated intersection. The type of queueing models we are dealing with are so-called polling models, models with multiple queues which share a common server, see e.g.~\cite{borst2018polling}. A study that is close to ours is \cite{boon2012delays}, where a similar type of traffic-light control is studied, except that there is no maximum length of the cycle. This might cause very long delays at e.g. lanes with a relatively small amount of traffic.
	
	Choices for parameters of actuated traffic lights thus have to be based on approximations or surrogate models. Examples of algorithms to choose those settings are based on e.g. reinforcement learning, model predictive control, backpressure, self-control, fuzzy logic and genetic algorithms, for which references are given in \cite[Section 1.5.2]{fleuren2017}. Other examples are using approximations for the mean queue length (see e.g.~\cite{weik2018traffic}  and \cite{timmerman2019platoon} for such a strategy applied to respectively railway systems and platoon forming algorithms for self-driving vehicles); the use of Markov decision theory, see e.g.~\cite{haijema2008source,cai2009adaptive}; direct use of (micro-)simulations, see e.g. \cite{zhang2010optimizing}; and so-called metamodels, an approach based on simulations where known information is exploited, see e.g.~\cite{osorio2013simulation,osorio2015computationally}. Two popular vehicle-actuated control mechanisms for networks are SCOOT \cite{hunt1981scoot} and SCATS \cite{sims1980sydney}, but due to their adaptive nature they provide less insight.
	
	In general, the approaches that are used in the literature obtain (close to) optimal settings with respect to some measure, but structural properties of such optimal settings are hard to derive. However, such results have recently been obtained for the Fixed-Cycle Traffic-Light (FCTL) queue for isolated intersections using a so-called heavy-traffic scaling, in which a particular relation between the cycle length and the time allocated to each lane is chosen \cite{boon2020inprogress}. In \cite{boon2020inprogress}, the authors are able to prove that when the capacity and demand are balanced (or scaled) in the right way, there exists an allocation of the access times for each of the lanes such that the probability of facing an empty queue at the end of the green period is strictly bigger than 0 even when the vehicle-to-capacity ratio approaches 1, see Section \ref{sec:theory} for more details. The heavy-traffic scaling is based on the Central Limit Theorem (CLT), see e.g.~\cite{grimmett2001probability}, intuitively meaning that the capacity for each lane should be chosen as the mean amount of ``work'' arriving during a cycle (time needed for all arriving vehicles to cross the intersection), where a variability hedge based on the square root of the variance of the amount of ``work'' is added. This approach has been applied in various settings and has proven its merits there, for more details see Section \ref{sec:theory}.
	
	In this paper, we extend the results for fixed green periods at isolated intersections to a fully actuated setting, so the length of the green period or access period will be random. We use the same scaling as for the FCTL queue in \cite{boon2020inprogress} with appropriate modifications and show that the same set of properties thus holds for more general settings of the access period. This enables us to gain understanding in the close to optimal settings for adaptive traffic lights, rather than just obtaining those values. Moreover, it is very easy to compute those values and to explain \emph{why} these settings are performing very well. These insights might also pave the way to obtain similar results for networks of intersections with an adaptive control, although such an extension is not straightforward.
	
	\subsection{Main contributions} \label{sec:main}
	
	Our first contribution is that we propose a new way of finding a good length of the access periods at intersections. We extend the results of \cite{boon2020inprogress} to a more general distribution of the length of the access period. Instead of a fixed length of the access period, we allow for an actuated control, thus having randomness and dependencies among different cycles in the access period.
	
	Moreover, our approach enables us to gain insights into close to optimal settings for adaptive traffic lights. Instead of the need for difficult optimization schemes or computationally expensive simulations, we are able to find the (close to) optimal settings on the basis of one-line calculations. Another advantage is that our scheme is easy to explain, which adds to its practical value.
	
	\subsection{Organization of the paper} \label{sec:organ}
	
	In Section~\ref{sec:model} we present a detailed description of the model we consider in this paper. Section~\ref{sec:theory} is devoted to sketching the theoretical background from which we take our inspiration. In Section~\ref{sec:sim} we present simulation results that provide valuable insights and we wrap up in Section~\ref{sec:con} with some conclusions and topics for further research.
	
	\section{MODEL DESCRIPTION} \label{sec:model}
	
	As mentioned before, we focus on isolated intersections, with a certain number of legs, $N$, leading towards it, which we number from $i=1,...,N$. We number the phases from $j=1,...,M$ and each phase $J_{j}$ satisfies $J_j \subseteq \{1,...,N\}$. We assume that each phase consists of one or more non-conflicting traffic flows, so that each lane in the same phase can receive access to the intersection at the same time. We also assume that each lane belongs to at least one phase.
	
	We will model such an intersection as a queueing model, in order to apply and extend the machinery developed in \cite{boon2020inprogress}. A cycle is a repeating sequence of traffic signal phases. For each lane, we divide this cycle into slots of fixed length, as in \cite{boon2020inprogress}, but instead of assuming a fixed length for the access period (the time during which vehicles are allowed to cross) for each lane as in the FCTL queue, we assume a vehicle-actuated control for each of the access periods in a cycle. The actuation mechanism is as follows: when all lanes in a phase do not have any vehicles in the queue anymore, the cycle immediately continues with the next phase or we switch to the next phase after a fixed maximum time of the queue is not yet dissolved. We assume that a lane remains empty as soon as a lane gets empty, as is done in \cite{boon2012delays}. This makes sense, as when the queue is dissolved during the access period, a new queue does not build during the remaining access period. We choose the access times as follows: initially, we start with the maximum length of the cycle, denoted with $c$, based on which we allocate the access times to each of the lanes. Any remaining part of $c$ is assumed to be a ``red'' time for all phases (which scales linearly with $c$), see also Remark \ref{rem:growing_red}.
	
	We introduce some further notation: $\lambda_{i}$ is the arrival rate of vehicles at lane $i$ in a single slot and the standard deviation of the number of arrivals at lane $i$ in each slot is denoted with $\sigma_i$. We assume that the number of arrivals during any slot for any lane are identically distributed and behave independently of one another. Further, with $g_{i,c}$ we denote the access period allocated to lane $i$ when the cycle length is $c$ and with $\beta_i$ we denote positive numbers, which might depend on the lane $i$. One could use this $\beta_i$ as a control parameter, with which we are able to steer the performance of the lanes: a higher $\beta_i$ implies a longer access period.
	
	\begin{remark}\label{rem:growing_red}
		In most settings, the sum of times during which no vehicles are allowed to cross the intersection, the all-red time, will be fixed in length. In order to show the type of properties we are after, we however assume that the all-red time increases with the cycle length. When letting go of this assumption, we stress that the performance  will \emph{improve}, while already having very good performance in our setting, as we will show in Section \ref{sec:sim}. There we also briefly comment on the implications when the all-red time is fixed.
	\end{remark}
	
	\begin{remark}
		The inspiration for this model is current-day traffic. However, exactly analogous results hold for vehicles that are autonomous, using methods from  \cite{timmerman2019platoon}. More examples of intersection access algorithms can be found in \cite{timmerman2019platoon}.
	\end{remark}
	
	\section{THEORETICAL BACKGROUND} \label{sec:theory}
	
	As said, we apply the ideas in \cite{boon2020inprogress}. First of all, we require stability of the underlying queueing model that we study, or equivalently: the vehicle-to-capacity ratio should be less than 1. This boils down to the requirement
	\begin{equation*}\label{eq:stability}
	g_{i,c} > \lambda_i c,
	\end{equation*}
	which intuitively makes sense: the capacity at lane $i$ (the left-hand side) should be higher than the average number of arrivals during a cycle (the right-hand side). This is a necessary and sufficient condition as it also is for the FCTL queue, see e.g. \cite{van2006delay}. When heavily loaded, the actuated control will behave similarly to the FCTL queue and therefore the stability condition is the same.
	
	We propose a new way to allocate the access times to lane $i$, where we exploit the heavy-traffic scaling result obtained in \cite{boon2020inprogress},
	\begin{equation}\label{eq:scaling}
	g_{i,c} = \lambda_{i}c+\beta_i\sigma_{i}\sqrt{c},
	\end{equation}
	where we note that $g_{i,c}$ might have to be rounded up (to ensure stability), as $g_{i,c}$ is a \emph{number} of slots. The scaling thus relates to finding the right access time or scale of the capacity for lane $i$, based on the maximum cycle length $c$. Moreover, when $g_{i,c}$ is chosen as in \eqref{eq:scaling}, the vehicle-to-capacity for each lane converges to 1 when $c$ converges to infinity. Even though the scaling rule \eqref{eq:scaling} is meant for the case when $c$ gets large and when we are operating close to oversaturation, we stress that also for small $c$ good performance is obtained, see Section \ref{sec:sim}. This scaling rule \eqref{eq:scaling} has been applied in many settings and in various guises. It yields very desirable properties in many respects: the limiting process, when $c$ grows large, is generally a well-understood process with good system performance. Examples (in our setting) are that the probability of an empty queue at the end of the access period is strictly between 0 and 1, instead of converging to 0, and that the mean queue length for each lane (which we measure in \emph{number of vehicles}) at the end of the access period scales with the cycle length $c$ as $\sqrt{c}$. Note that the additional capacity needed, compared to the minimum of $\lambda_ic$ required for stability, is only of order $\sqrt{c}$, so it increases very slowly compared to the leading order term.
	
	Moreover, the limiting process usually yields good and easy-to-use approximations for a plethora of complicated systems, which can then be used for further purposes like optimization of certain performance characteristics. Examples of the latter can be found in e.g. call centers \cite{borst2004dimensioning}, communication systems \cite{tan2012heavy} and traffic engineering \cite{boon2020inprogress}. In \cite{borst2004dimensioning}, it is shown that an approximation based on the limiting process yields very good results even when ``far from the limit'' (this relates to small $c$ in actuated-access control): an optimization based on the approximation, yields an optimal allocation for many parameter settings.
	In \cite{boon2020inprogress} a similar approach is taken and qualitatively similar results are obtained. Based on these observations, we expect that the scaling that we propose results in similar \emph{optimality} results for the actuated access control, yet we do not study this in-depth. For further background on the scaling rule and more references, we refer to the tutorial and review paper \cite{van2019economies}.
	
	The scaling as in \eqref{eq:scaling} is inspired by the CLT, a fundamental tool in probability theory. A sum of random variables (under some conditions on independence and similarity) can be scaled by subtracting the mean and dividing by the standard deviation after which the distribution of the sum can be approximated by the normal distribution. Even though we scale the capacity for each queue (the length of the access period) and the demand (the cars aiming to pass the intersection), we have the same structure as in the CLT. The demand is simply a sum of random variables and the capacity is just the appropriate scaling: the mean of the demand is $\lambda_{i}c$ and to ensure stability, we add a small term, namely $\beta_i\sigma_{i}\sqrt{c}$. Then, after multiplying with a factor $1/(\sigma_{i}\sqrt{c})$, and letting $c$ tend to infinity, we see that the right approximation is a normal distribution with mean $-\beta_i$ (ensuring stability) and variance 1. This is also the approach taken in \cite{boon2020inprogress}, where this intuition is shown to be the exact outcome of the scaling.
	
	Due to the vehicle-actuated control considered in this paper, we are not able to show analytical results. However, our simulation results indicate that we are in the right window: the probability of an empty queue at the end of the access period is strictly between 0 and 1 and the queue length just before switching to the next queue is of order $\sqrt{c}$. The difficulty when considering actuated-access control is that dependencies between queues carry over: when a queue empties early in this cycle, the other queues are likely to be short as well. For this dependence we cannot account in exact computations. From the literature on polling systems, it is known that these types of actuated queueing models offer little or no hope on an exact solution, see e.g. \cite{borst2018polling}.
	
	\section{RESULTS} \label{sec:sim}
	
	In this section, we show the desirable properties of an actuated traffic control with a scaling rule between demand and capacity as in \eqref{eq:scaling}. We employ (discrete-event) simulations in order to gain those insights. We consider various settings and discuss the differences, similarities and relate them to the results in \cite{boon2020inprogress}. We also validate (part of) our results with the microscopic traffic simulator SUMO \cite{lopez2018microscopic}, which captures more realistic features such as interactions between vehicles and acceleration/deceleration.
	
	\subsection{Single-lane access control}
	
	\subsubsection{Example 1a}
	
	This example consists of four lanes, so $N=4$, where each lane has its own dedicated phase, i.e. cars from only one lane are allowed to cross the intersection, so $J_j = \{j\}$ for $j=1,...,4$. We assume that all vehicles are going straight and that the arrivals per time slot are Poisson distributed, with means $\lambda_i = i/11$, for $i=1,...,N$. In this way, we will be able to assign appropriate  $g_{i,c}$ for any $c$ sufficiently big to ensure stability. We choose $\beta_{i} = 1/10$ and if \eqref{eq:scaling} results in non-integer values for $g_{i,c}$, we round them up to the nearest integer.
	
\begin{figure*}[ht]
		\centering
		\begin{minipage}[b]{0.45\linewidth}
			\includegraphics[width=\textwidth]{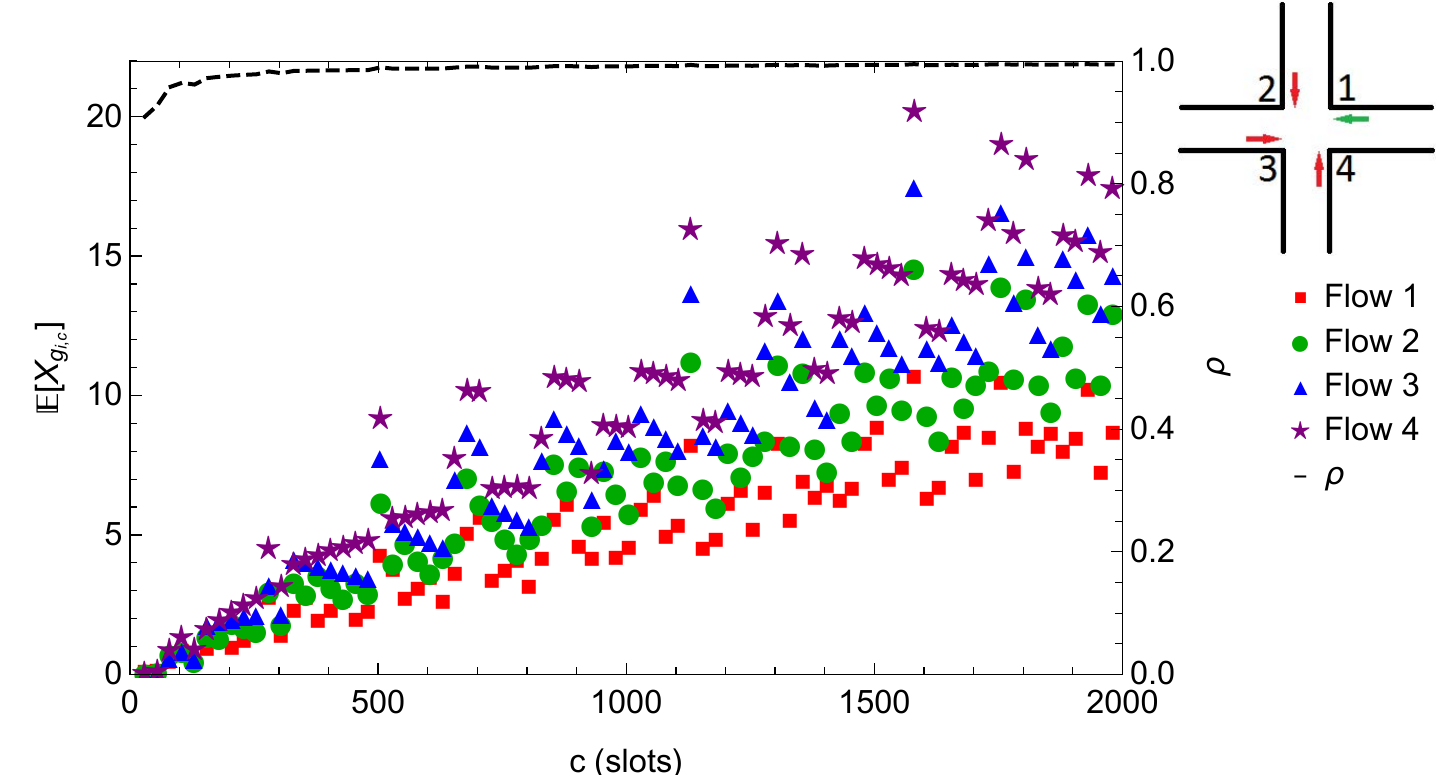}
			\caption{The simulated expected queue length at the end of the access period for the four traffic flows in Example 1a and the vehicle-to-capacity ratio $\rho$ on the right axis (dashed line).}
			\label{f:exgEx1klim}
		\end{minipage}
		\hspace{1 cm}
		\begin{minipage}[b]{0.45\linewidth}
			\includegraphics[width=\textwidth]{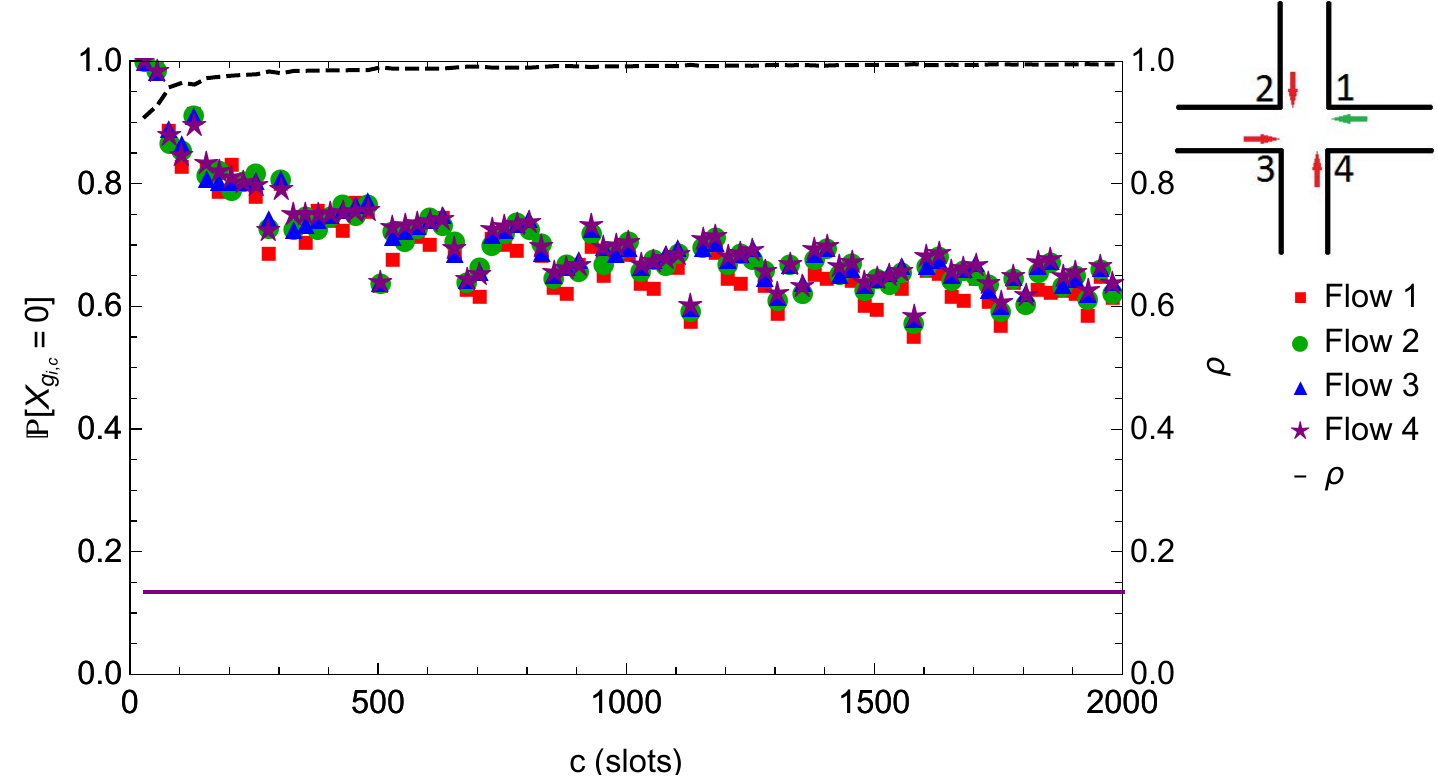}
			\caption{The simulated probability of an empty queue at the end of the access period for the four traffic flows in Example 1a, the limiting value of that probability in the fixed access control (solid lines) and the vehicle-to-capacity ratio $\rho$ on the right axis (dashed line).}
			\label{f:px0Ex1klim}
		\end{minipage}
	\end{figure*}

	To investigate the influence of the cycle length $c$ on the mean queue length at the end of the access period, $\mathbb{E}[X_{g_{i,c}}]$, and the probability of having an empty queue at the end of the access period, $\mathbb{P}(X_{g_{i,c}}=0)$, we perform efficient discrete-event simulations. We perform 8 independent runs with a length of $10^7$ cycles in order to reduce simulation variability and obtain accurate simulation results (we also take this number of cycles and runs for the other examples we present unless otherwise specified). The results are shown in Fig. \ref{f:exgEx1klim} and \ref{f:px0Ex1klim}. Note that the dashed black line (as in the other figures) represents the weighted sum of the vehicle-to-capacity ratios of each lane separately, $\rho$, with its value on the right axis.
	
	\begin{figure*}[ht]
		\centering
		\begin{minipage}[b]{0.45\linewidth}
			\includegraphics[width=\textwidth]{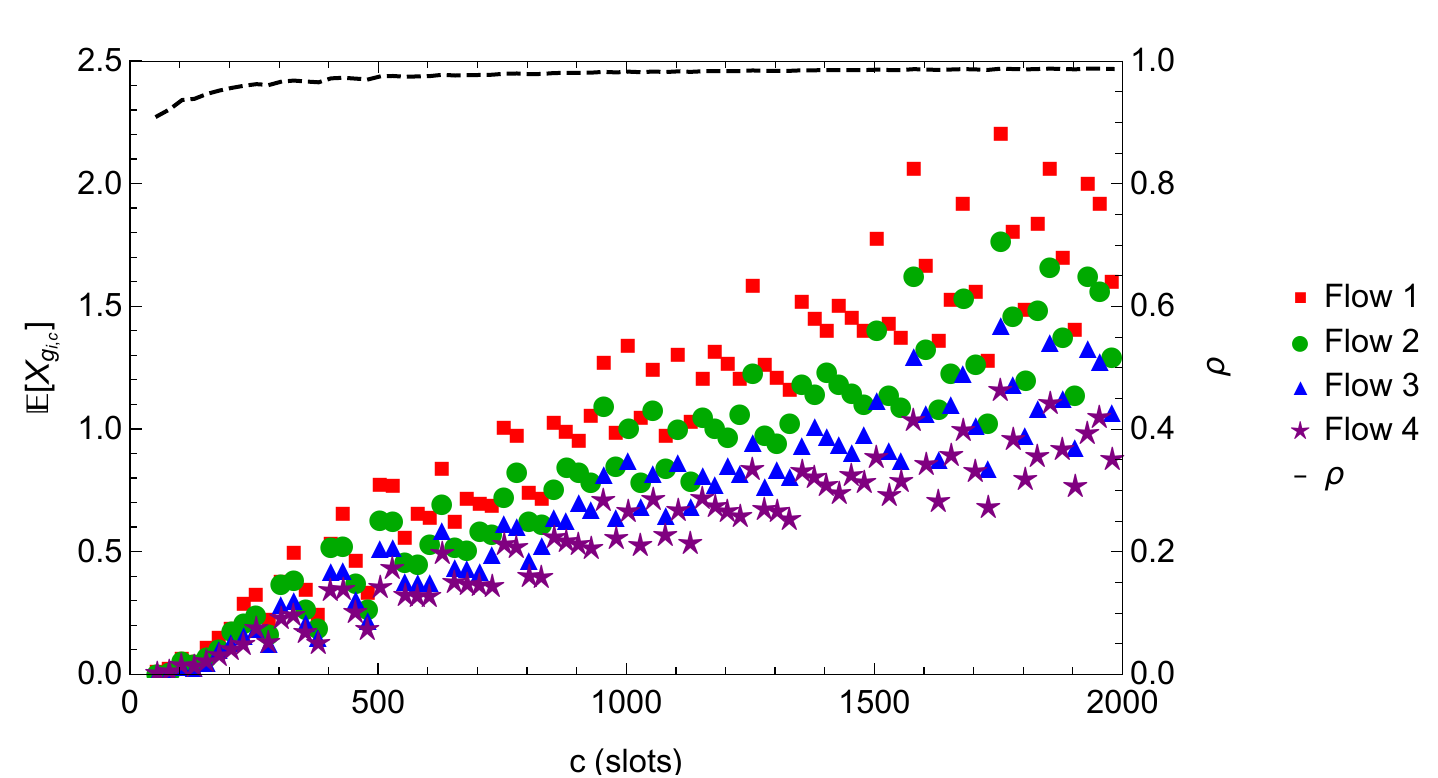}
			\caption{The simulated expected queue length at the end of the access period for the four traffic flows in Example 1b and the vehicle-to-capacity ratio $\rho$ on the right axis (dashed line).}
			\label{f:exgEx1klimas}
		\end{minipage}
		\hspace{1 cm}
		\begin{minipage}[b]{0.45\linewidth}
			\includegraphics[width=\textwidth]{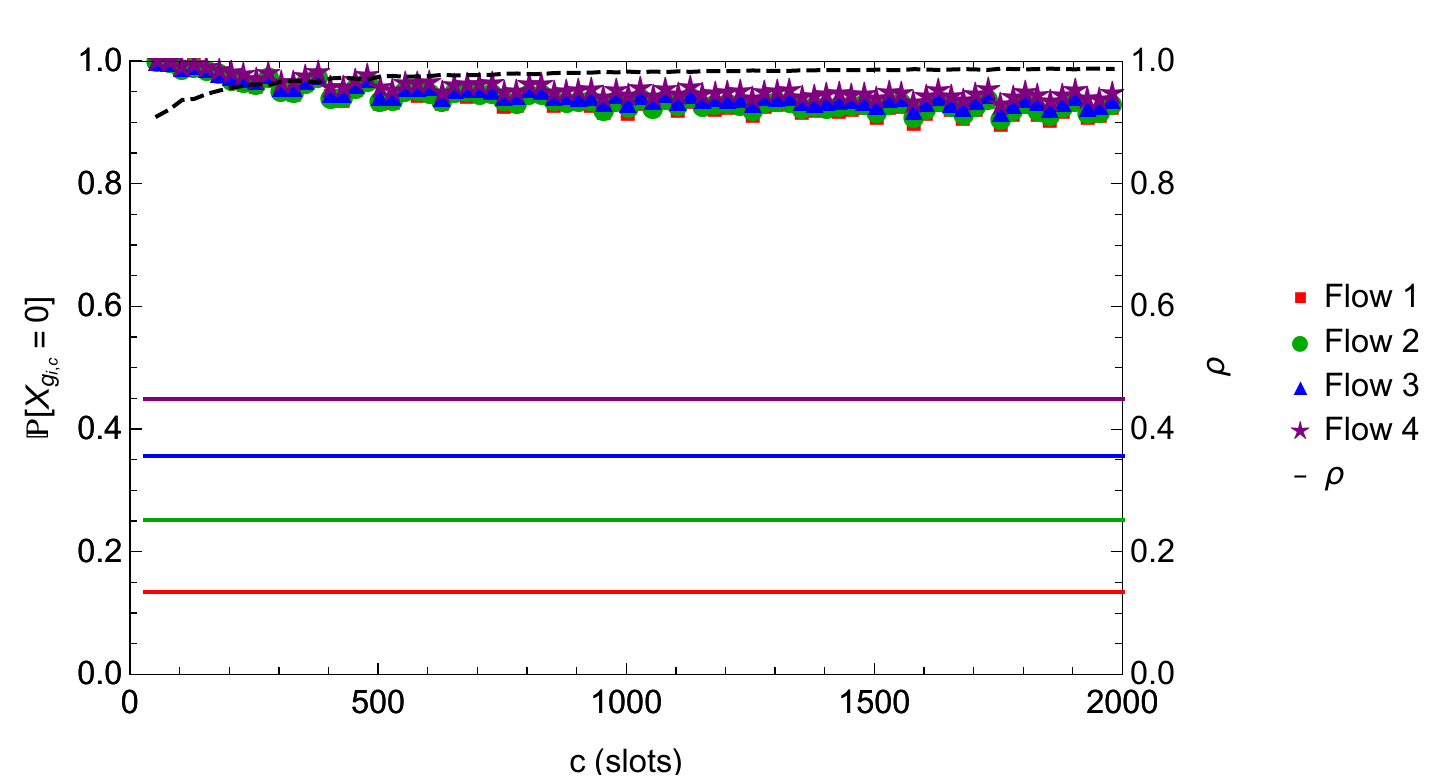}
			\caption{The simulated probability of an empty queue at the end of the access period for the four traffic flows in Example 1b, the limiting value of that probability in the fixed access control (solid lines) and the vehicle-to-capacity ratio $\rho$ on the right axis (dashed line).}
			\label{f:px0Ex1klimas}
		\end{minipage}
	\end{figure*}
	\begin{figure}[h]
		\centering
		\includegraphics[width=0.90\linewidth]{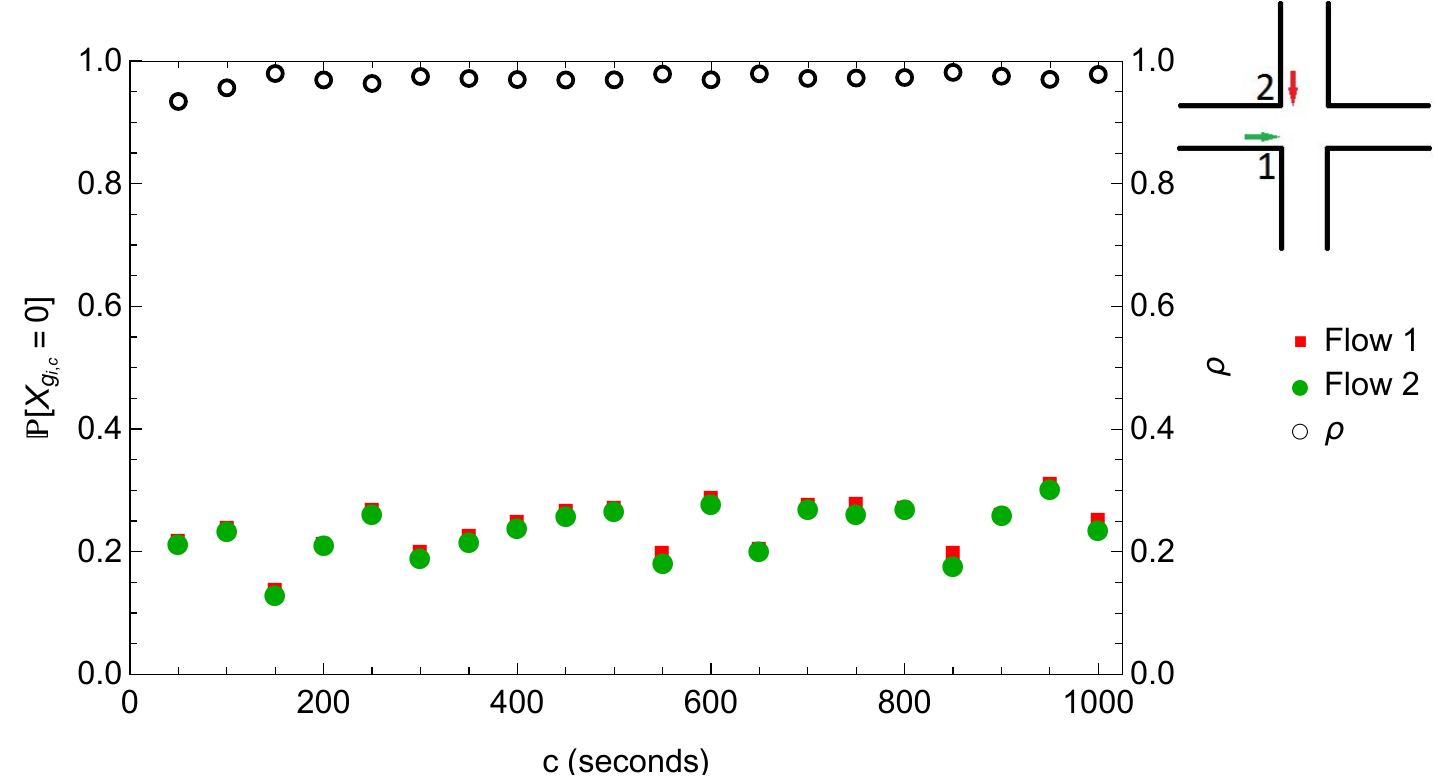}
		\caption{The simulated probability of an empty queue at the end of the access period for each of the two traffic flows in the SUMO Example and the black circles represent the vehicle-to-capacity ratio $\rho$ on the right axis.}
		\label{f:sumo}
	\end{figure}
	
	The mean queue length at the end of the access period scales with $\sqrt{c}$, which is shown by the results in Fig. \ref{f:exgEx1klim}. The mean queue length grows as a constant times $\sqrt{c}$, as after dividing by $\sqrt{c}$, the mean queue length seems to converge to a constant (modulo the rounding effect of the $g_{i,c}$ and the simulation uncertainty). The higher the load on a lane, the higher the limiting constant as is the case in Fig. \ref{f:exgEx1klim}. There is no influence of $\beta_i$ visible in this example, as the value of $\beta_i$ is the same for all lanes. The vehicle-to-capacity ratio in Fig. \ref{f:exgEx1klim} is well above 0.9 for all values of $c$ and often above 0.99, which makes the low mean queue lengths at the end of the access period quite remarkable.
	
	The predicted behavior of the probability that a queue is empty at the end of the access period is observed in Fig. \ref{f:px0Ex1klim}: the probability converges to a value between 0 and 1. This value is independent of the load on the lane, as is the case for the FCTL queue under the same type of scaling \cite{boon2020inprogress}.
	
	The actuated access control mechanism clearly outperforms the fixed control mechanism. This is visible in Fig. \ref{f:px0Ex1klim}, as the limiting probability for the fixed control scheme is below 0.2 (this value is in accordance with \cite{boon2020inprogress}), whereas the simulated probabilities for the actuated access control are well above 0.5. The expected queue length for the fixed access control is different for each of the lanes, as is the case for the actuated control setting in Fig. \ref{f:exgEx1klim}. The mean queue length for the FCTL queue would be much higher than the values in Fig. \ref{f:exgEx1klim}, which is why we did not plot these results.

	\begin{figure*}
		\centering
		\begin{minipage}[b]{0.45\linewidth}
			\includegraphics[width=\textwidth]{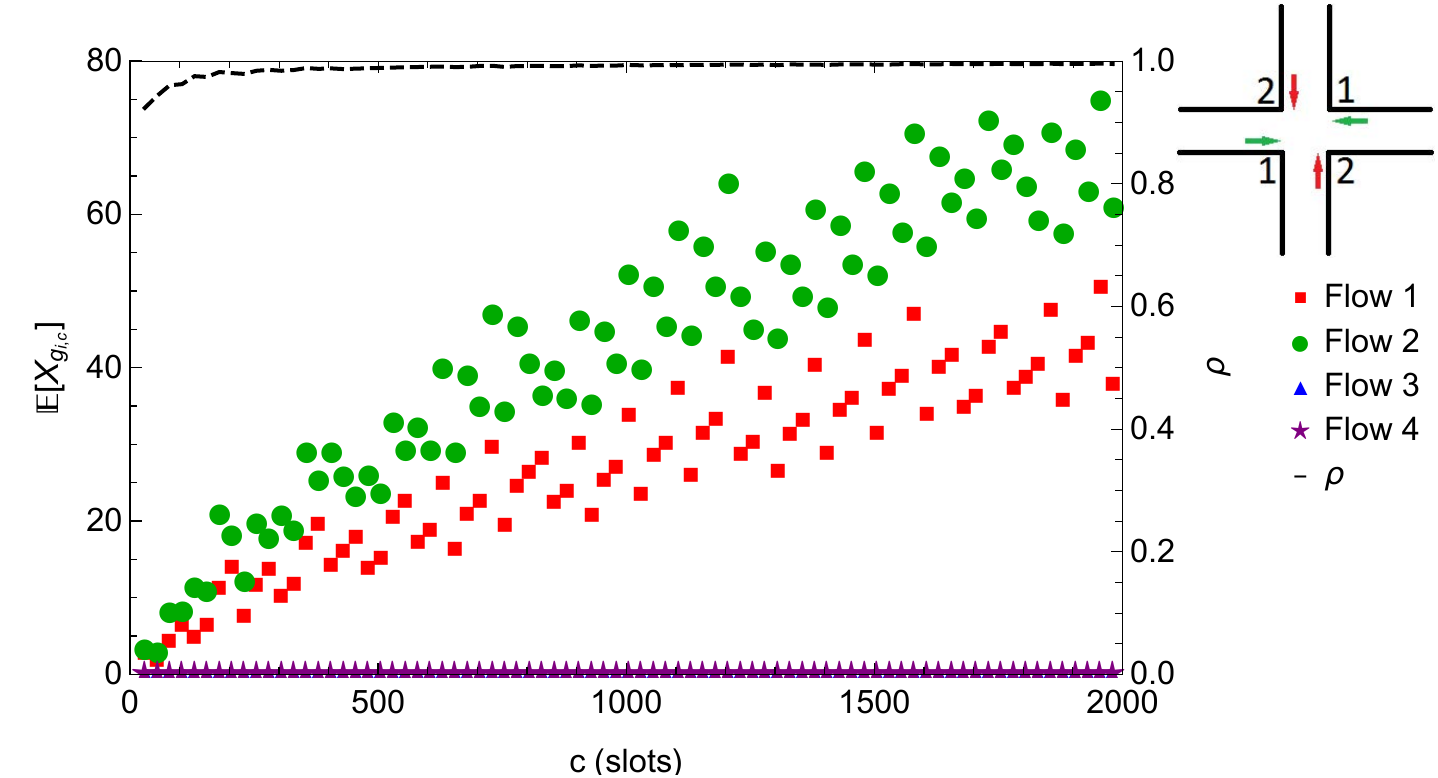}
			\caption{The simulated expected queue length at the end of the access period for the four traffic flows in Example 2a and the vehicle-to-capacity ratio $\rho$ on the right axis (dashed line).}
			\label{f:exgEx2klimas}
		\end{minipage}
		\hspace{1 cm}
		\begin{minipage}[b]{0.45\linewidth}
			\includegraphics[width=\textwidth]{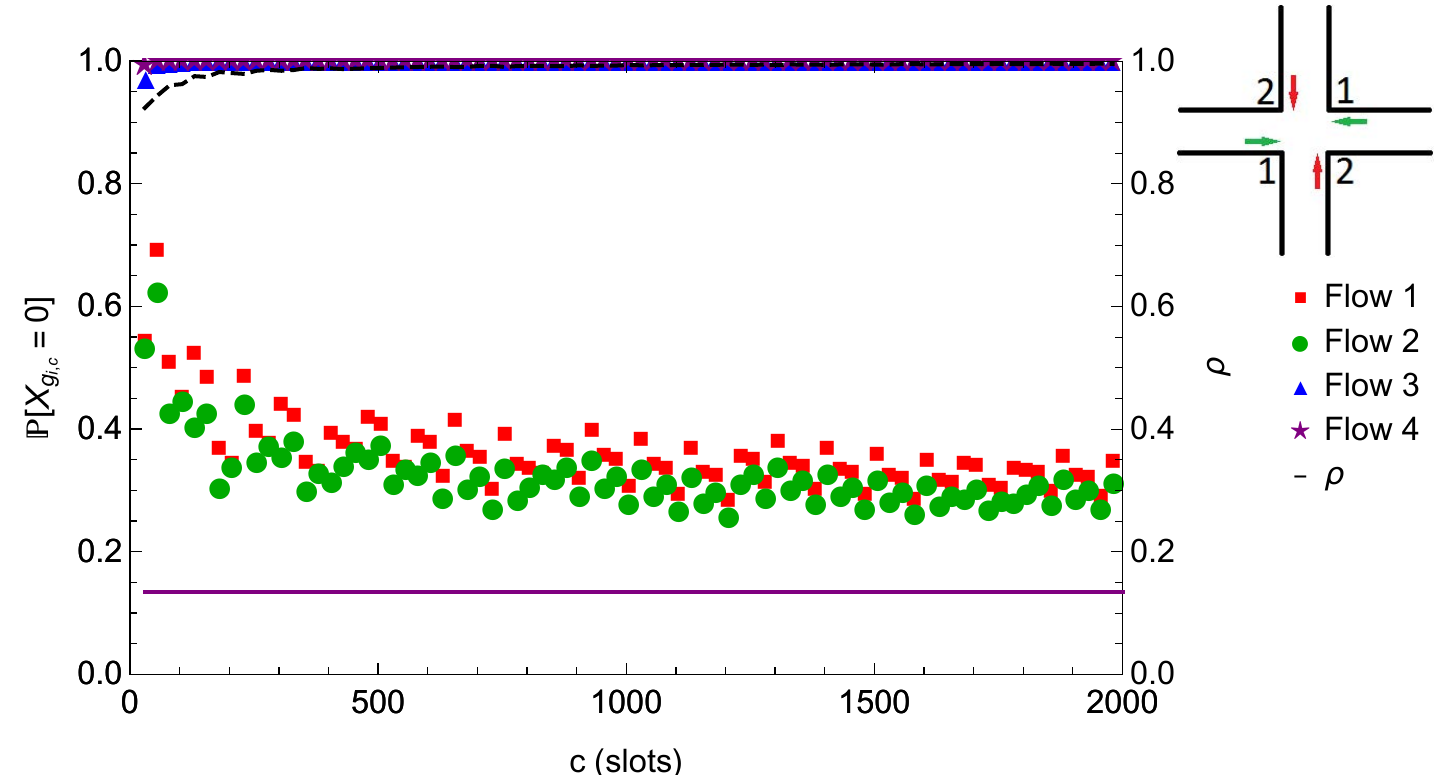}
			\caption{The simulated probability of an empty queue at the end of the access period for the four traffic flows in Example 2a, the limiting value of that probability in the fixed access control (solid lines)  and the vehicle-to-capacity ratio $\rho$ on the right axis (dashed line).}
			\label{f:px0Ex2klimas}
		\end{minipage}
	\end{figure*}
	\begin{figure*}
		\centering
		\begin{minipage}[b]{0.45\linewidth}
			\includegraphics[width=\textwidth]{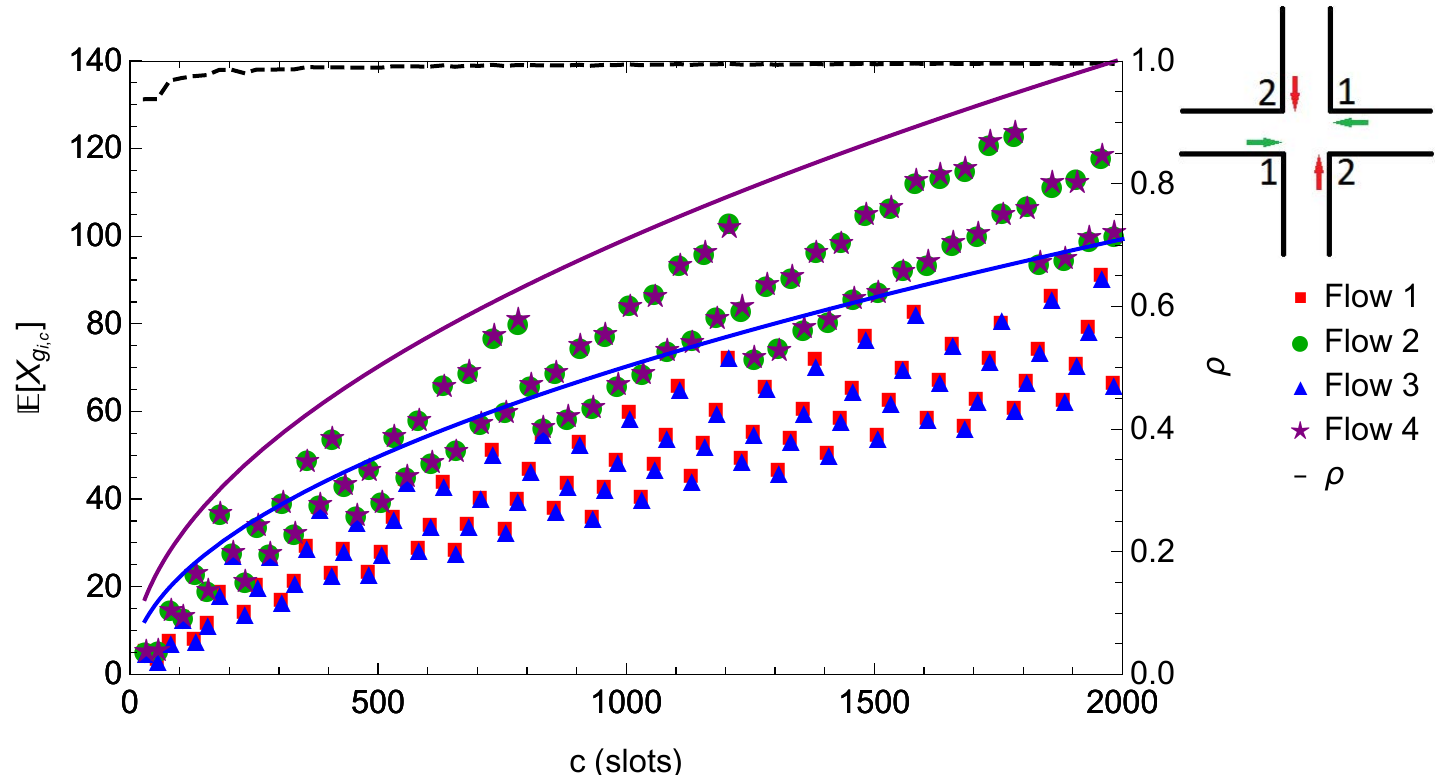}
			\caption{The simulated expected queue length at the end of the access period for the four traffic flows in Example 2b, the limiting expected queue length for the FCTL queue (solid lines) and the vehicle-to-capacity ratio $\rho$ on the right axis (dashed line).}
			\label{f:exgEx2klim}
		\end{minipage}
		\hspace{1 cm}
		\begin{minipage}[b]{0.45\linewidth}
			\includegraphics[width=\textwidth]{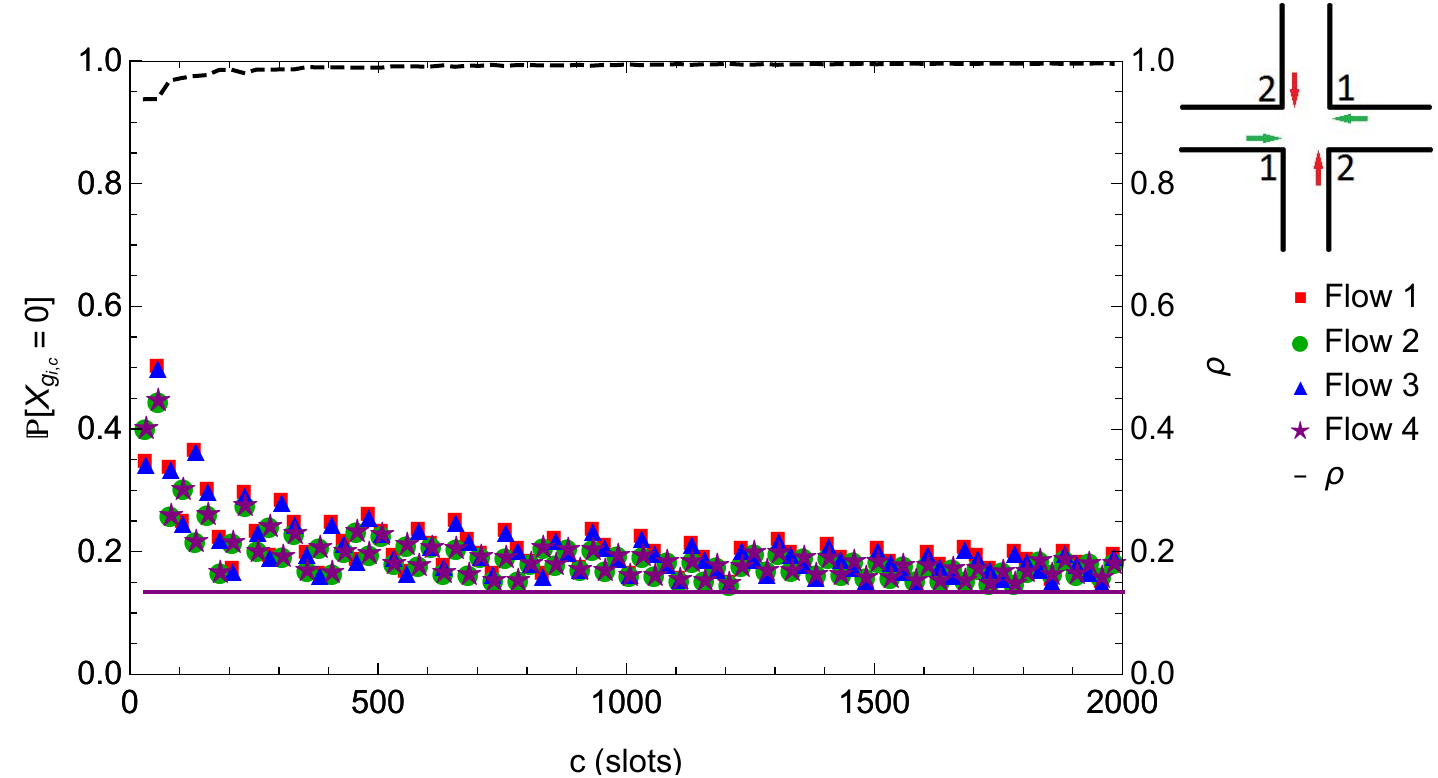}
			\caption{The simulated probability of an empty queue at the end of the access period for the four traffic flows in Example 2b, the limiting value of that probability in the fixed access control (solid lines) and the vehicle-to-capacity ratio $\rho$ on the right axis (dashed line).}
			\label{f:px0Ex2klim}
		\end{minipage}
	\end{figure*}
	
	If we would assume a fixed length for the all-red period and actuated control, Fig. \ref{f:exgEx1klim} and \ref{f:px0Ex1klim} would change considerably. The mean queue length at the end of the access period converges to 0 and the probability of an empty queue at that same moment converges to 1. This shows that, as long as the vehicle-to-capacity ratio is below 1 and under some independence assumptions on the arrivals, an actuated traffic control can lead to those desirable properties.
	
	\subsubsection{Example 1b}
	
	In Fig. \ref{f:exgEx1klimas} and \ref{f:px0Ex1klimas} we have adapted the values of $\lambda_i$ and $\beta_i$. We choose $\lambda_i = 5/22$ and $\beta_i=i/10$. Qualitatively, we observe the same behavior, yet some interesting differences are also present. The mean queue length at the end of the access period depends on the value of $\beta_i$, which makes sense: a higher $\beta_i$ results in a longer maximal access period, and thus in a smaller queue length.
	
	Surprisingly, the probability of an empty queue at the end of the access period seems to converge to the same value for each of the lanes, even though the values of $\beta_i$ differ. In the fixed access control setting there is a differentiation, see Fig. \ref{f:px0Ex1klimas}. It might be that an empty queue implies an early switch to the next phase. This is then more likely to result in empty queues at the end of the access period in that phase, because the vehicles had a smaller time to accumulate on these lanes. This effect seems to strengthen over cycles and to cause the probability to be the same for each of the lanes.
	
	\subsubsection{SUMO Example}
	
	As a proof of concept, we also present an example performed in the microscopic traffic simulator SUMO employing the so-called vehicle-actuated control of SUMO based on time gaps. Our purpose is to show that also in this simulator, which is generally considered to be excellent in capturing real-world traffic dynamics, we are able to define an actuated control with the desirable properties as are obtained in the other examples. For simplicity, we assume that we have two lanes, so $N=2$. As arrival distribution we choose a Bernoulli distribution with parameter $\lambda_{i}=3/20$ (these arrivals correspond to a single simulation step in SUMO) and $\beta_i=1/10$. We choose to do a single simulation run of 3,600,000 steps, because this gives results that are (more than) sufficiently accurate for our purposes.
	
	One of the main difficulties in this example is to define a slot, as the period between departures is not constant. Partly because of this, it is also difficult to compute the vehicle-to-capacity ratio and to determine if we are close to oversaturation. We obtain a good measure for this ratio by dividing the average effective vehicle access time per cycle by the maximum specified access period. This ratio is close to one, because we are close to oversaturation.
	
	In Fig. \ref{f:sumo} we see the probability of an empty queue at the end of the access period (as determined by the actuated control mechanism in SUMO) and the vehicle-to-capacity ratio. Qualitatively, we observe similar behavior as in Examples 1a and 1b. The vehicle-to-capacity ratio approaches 1 quickly, yet the probability of an empty queue remains between 0 and 1, which shows that our discrete-event simulations are able to capture the essential queueing behavior of vehicles at intersections with an actuated control sufficiently well.
	
	\subsection{Multiple-lane access control}
	
	\subsubsection{Example 2a}
	
	Also in this example we assume that the intersection has four legs and only straight-going traffic, yet here we combine the two opposing non-conflicting lanes in a single phase, i.e. $J_{1} =\{1,3\}$ and $J_{2} = \{2,4\}$. This allows for a higher load on the intersection, as twice as many vehicles are allowed to depart at the same time in comparison with the Examples 1a and 1b. We choose $\lambda_1=1/4$, $\lambda_2=1/2$, $\lambda_3=3/20$ and $\lambda_4=3/10$, all arrival distributions to be Poisson and $\beta_i=1/10$. We present results for the mean queue length and the probability of an empty queue both at the end of the access period, see Fig. \ref{f:exgEx2klimas} and \ref{f:px0Ex2klimas}.
	
	Lanes 1 and 2, the lanes with the highest load, show similar behavior as in Examples 1a and 1b both for the actuated and the fixed control. However, different behavior is observed for lanes 3 and 4. The access period is too long, because the length of the access period is dominated by lanes 1 and 2 that face more traffic. This implies that the queue is (very often) empty at the end of the access period for both lanes 3 and 4 (the purple stars are on top of the blue triangles), as can be observed in Fig. \ref{f:px0Ex2klimas}. 
	
	\addtolength{\textheight}{-0cm}
	
	\subsubsection{Example 2b}
	
	The last example that we discuss is the same as Example 2a, except that $\lambda_3 = 1/4$ and $\lambda_4 = 1/2$. In this example, the load on the lanes in each of the phases is the same. In Fig. \ref{f:exgEx2klim} and \ref{f:px0Ex2klim} we see that both lanes inside a phase behave similarly and we observe the same desirable properties as in the other examples. 
	
	When comparing Fig. \ref{f:exgEx2klimas} and \ref{f:exgEx2klim}, we see that the limiting value of the mean queue length is considerably higher in Example 2b. This is the result of having longer access periods on average for both phases, as we only switch to the next phase when \emph{both} queues are empty, while both queues are on average equal in length.  So, usually we switch later to the next phase in Example 2b, which causes the queues at other lanes to be longer, resulting in a higher mean queue length. It even turns out that the mean queue lengths in Fig.~\ref{f:exgEx2klim} are close to a corresponding FCTL queue, indicating that the access periods are often (almost) of maximum length. The same intuition seems to hold for the decrease in the probability of an empty queue, see Fig. \ref{f:px0Ex2klimas} and \ref{f:px0Ex2klim}.
	
	Examples 2a and 2b do not immediately indicate that a convenient setup of each phase is one in which each of the lanes has more or less the same load. The lane with the highest load is dominating the length of the phase in Example 2a (which is favorable), but some capacity is ``lost'' for the lanes with a lower load in that phase. When there is a queue, the outflow is a single car per slot. However, if there is no queue at a lane, we have a lower outflow equal to the arrival rate, which is strictly smaller than 1. On the other hand, the longer access periods in Example 2b cause some negative effects as well, due to longer time periods in which the queues at other lanes can accumulate. Based on this, it is not clear which is the best option. Note that a direct comparison of the mean queue lengths is not fair, as the total load on the intersection is not the same in the two examples.
	
	\section{CONCLUSION} \label{sec:con}
	
	We have shown with the aid of simulation that desirable properties are achievable for actuated traffic control of isolated intersections when using a scaling rule such as \eqref{eq:scaling}. Those properties are similar to the ones established for the FCTL queue in \cite{boon2020inprogress}. We have investigated several setups and in each of those, we have observed those desirable properties. A prominent one is that the limiting probability of an empty queue at the end of the access period is strictly between 0 and 1. We also observed this in our simulation experiment that we performed in SUMO, indicating that our results are qualitatively reliable for real traffic (remember that the other simulations are discrete-event simulations). Another desirable property is that the mean queue length at the end of the access period grows only with order $\sqrt{c}$.
	It would be interesting to check whether these desirable properties carry over to a network of intersections. Since an analytic network study is not straightforward (see e.g. \cite{boon2018networks}), the intuition obtained from this study will be an important starting point for this future study.

	
	\section*{ACKNOWLEDGMENTS}
	
	We would like to thank Johan van Leeuwaarden and Onno Boxma for helpful discussions.
	
	\bibliographystyle{IEEEtran}
	\bibliography{ref}

\end{document}